\documentclass[aps,prb,showpacs,twocolumn,amssymb]{revtex4}

\usepackage{psfig}
\usepackage{amsmath}

\input epsf

\def\prb{Phys. Rev. B }
\def\prl{Phys. Rev. Lett. }

\def\be{\begin{equation}}
\def\ee{\end{equation}}
\def\ba{\begin{eqnarray}}
\def\ea{\end{eqnarray}}

\begin{document}

\title{The Tomonaga-Luttinger Model and the Chern-Simons Theory for the Edges of Multi-layer
Fractional Quantum Hall Systems}

\author{D. Orgad}
\affiliation{Racah Institute of Physics, The Hebrew University, Jerusalem 91904, Israel}

\date{\today}

\begin{abstract}

Wen's chiral Tomonaga-Luttinger model for the edge of an $m$-layer quantum Hall system of total
filling factor $\nu=m/(pm\pm 1)$ with even $p$, is derived as a random-phase approximation of the
Chern-Simons theory for these states. The theory allows for a description of edges both in and out of
equilibrium, including their collective excitation spectrum and the tunneling exponent into the
edge. While the tunneling exponent is insensitive to the details of a $\nu=m/(pm + 1)$ edge, it
tends to decrease when a $\nu=m/(pm - 1)$ edge is taken out of equilibrium. The applicability of
the theory to fractional quantum Hall states in a single layer is discussed.

\end{abstract}


\maketitle

\section{Introduction}
\label{sec:intro}

The physics at the edges of quantum Hall systems has been the subject of both theoretical and
experimental research since the early days following the discovery of the integer quantum Hall
effect\cite{Halperin82}, and up to the present \cite{Chang03}. When the system exhibits the integer
or the fractional quantum Hall effect (FQHE) the edge attains a unique role as the only region in
the sample where gapless excitations exist. The low-energy physics associated with these edge
excitations is especially rich under the conditions of the FQHE. In a pioneering series of papers
\cite{Wen-papers} Wen has developed a theory for the edges of FQHE samples in terms of several
interacting chiral Tomonaga-Luttinger liquids (CTLL), and predicted a power-law
current-voltage characteristic for tunneling into such edges.

Subsequent tunneling experiments \cite{Chang96,Grayson98,Chang00,Hilke01} have found a non-ohmic
conductivity $I\sim V^\gamma$, with an exponent which approximately (and non universally\cite{Hilke01})
scales as $\gamma\sim 1/\nu$. However, with the exception
of filling factor $\nu=1/3$, this result is at odds with the predictions of Wen's theory. Moreover,
the experiments suggest that a single charged-mode CTLL model appears to apply to filling factors
in which the system is compressible and lacks the energy-gap needed to distinguish the edge as the
region where the low-energy physics resides. While various theories have been put forward
there is no consensus on the solution to the problem\cite{Levitov01}.

In order to better understand the conditions under which the CTLL model for the FQHE edge is
applicable it is desirable to test its predictions and study the way it arises from the underlying
theory using as many theoretical approaches as possible. Previous research aiming at achieving this
objective has included numerical calculations\cite{Palacios96,Mandal02,Wan05} as well as analytical
studies with Laughlin's wave-function\cite{Boyarsky04} and the Chern-Simons (CS)
theory\cite{Orgad97,Yu97} for the FQHE as their departure points. These studies have been concerned
predominantly with the Laughlin fractions $\nu=1/(p+1)$, with even $p$. Here we wish to extend our
previous treatment of Ref. \onlinecite{Orgad97} to multi-component FQHE systems. We concentrate on
$m$-layer quantum Hall systems with total filling factor $\nu=m/(pm\pm 1)$, with even $p$, whose
state is described by an $m$-component generalization of the Laughlin wave-function, as first
introduced by Halperin\cite{Halperin-m}. We also consider the issues which arise when trying to
apply a similar treatment to the Jain fractions, i.e. to the single-layer FQHE states with the same
filling factors.

Our plan is to obtain the one-dimensional CTLL theory for the edge as the low-energy limit of the
underlying CS theory for the two-dimensional system. The CS description, presented in Section
\ref{sec:CS}, follows closely the points of view taken by Jain\cite{Jain89} and Wen\cite{Wen95}. We
first transform the electrons to composite fermions which fill, for the special filling factors
under consideration, the lowest effective Landau level in each layer of the multi-layer system or
$m$ such levels in the single-layer problem. We then transform each of the layers (or in the case
of the Jain states, filled Landau levels) into a bosonic condensate. The study of the coupled
dynamics of these condensates constitutes the major part of the work.

We begin, in Section \ref{sec:MFAsol}, by considering the solutions to the CS mean-field equations.
We find that the solutions form an $m$-dimensional manifold parameterized by the conserved momenta
along the edge of the $m$ condensates, or equivalently, by the charges which they carry. Only a
small fraction of the manifold describes equilibrium configurations which minimize the energy for
a given total charge at the edge. These configurations are typically, but
not always, symmetrical with all the condensates sharing the same state. The remaining solutions
correspond to edges which are not in equilibrium and are generally asymmetrical. The conditions for
the realization of a particular solution are discussed.

Once the mean-field configurations are at hand we proceed to consider small oscillations about them
and obtain the edge excitations as the random-phase approximation (RPA) modes of the theory. We
identify $m$ RPA branches of edge excitations whose $k\rightarrow 0$ limit is composed of the edge
states of the $m$ types of vortex-like excitations that exist in the bulk. Out of these $m$ modes one
is the edge magnetoplasmon which arises from in-phase oscillations of the condensates while the
remaining posses acoustic dispersion, propagate with or against the magnetoplasmon [for
$\nu=m/(pm+1)$ and $\nu=m/(pm-1)$ respectively] and describe out of phase motion of the electron
liquids. While most characteristics of the edge modes do not depend on the detailed nature of the
underlying mean-field solution there are, however, some differences between the excitations of
symmetrical and asymmetrical configurations; for example the net charge which is involved in the
fluctuations. Consequently, we consider the two cases separately in Sections \ref{sec:Sym} and
\ref{sec:Asym}.

Finally, by using the edge modes to expand the deviations of the various fields from their
mean-field average configurations we recover the CTLL model for the edge. The expansion
coefficients become upon quantization the density operators in terms of which the model is
formulated. Their commutation algebra as well as their Hamiltonian are deduced from the quadratic
expansion of the CS theory in the deviations. By inverting the CS transformation we also obtain the
bosonization formula which expresses the original electronic operator at the edge in terms of the
density operators. These results can be used to calculate the tunneling exponent into the edge. In
accord with the results of Wen\cite{Wen95}, we find that the exponent is insensitive to the details
of the edge configuration or the value of $m$ when $\nu=m/(pm+1)$. On the other hand, it tends to
decrease when a $\nu=m/(pm-1)$ edge is taken out of equilibrium.

\section{The Chern-Simons Theory}
\label{sec:CS}

In the following we consider a semi-infinite sample subjected to a constant magnetic field
$B\hat{\bf{z}}$, with an edge defined by an external potential $A_{0}(x)$ and periodic boundary
conditions over length $L$ in the $y$ direction. The electronic spin is assumed to be polarized by
the magnetic field and will be left out of the analysis.

We are interested in studying a particular correlated state of the $m$-layer system in which each
of the layers is in a $\nu=1/(pm \pm 1)$ state with strong inter-layer correlations. An appropriate
wave-function to describe such a system was introduced by Halperin\cite{Halperin-m}. For a
two-layer system it is of the form

\ba
\label{H-wf}
\nonumber
\Psi&\propto&\prod_{\alpha<\beta}(z_{1\alpha}-z_{1\beta})^{p\pm 1}
\prod_{\alpha<\beta}(z_{2\alpha}-z_{2\beta})^{p\pm 1} \\
&\times&\prod_{\alpha,\beta}(z_{1\alpha}-z_{2\beta})^{p}
\,e^{-\frac{1}{4}\sum_{I,\alpha}|z_{I\alpha}|^2} \; ,
\ea
where $z_{I\alpha}$ is the complex coordinate (in units of the magnetic length $l=\sqrt{\hbar
c/eB}$) of electron $\alpha$ in layer $I$.

An alternative field theoretical description of the same states may be achieved via the CS
transformation from the original electronic operators on the $m$ layers, $\Psi_I$, to new operators,
$\psi_I$, describing composite fermions made out of $p$ flux quanta $\phi_0=hc/e$ attached to each
electron\cite{Jain89}
\be
\label{cstransform1}
\psi_{I}({\bf r})=e^{i\Lambda({\bf r})}\Psi_I({\bf
r}) \;,\;\;\;\;\; \Lambda({\bf r})=\int d^2r' f_{1}({\bf r-r'})\rho({\bf r'}) \; ,
\ee
where $\rho({\bf r})=\sum_I\Psi^{\dagger}_I({\bf r})\Psi_I({\bf r})=\sum_I\psi_{I}^{\dagger}({\bf r})
\psi_{I}({\bf r})$ is the total density. The new fields are fermionic provided $f_{1}({\bf
r-r'})=f_{1}({\bf r'-r})+p\pi$ and $p$ is even. The CS transformation introduces additional terms
in the equation of motion of the $\psi_{I}({\bf r})$ which couple to them as vector and scalar
potentials
\begin{eqnarray}
\label{axy1}
&& {\bf a}({\bf r})=-\frac{\hbar c}{e}\int d^2r' {\bf F}_{1}({\bf r-r'})\rho({\bf r'}) \; , \\
\label{a01} && a_{0}({\bf r})=-\frac{\hbar}{e}\int d^2r' {\bf F}_{1}({\bf r-r'})\rho({\bf r'})\mbox
{\boldmath $v$}({\bf r'}) \; ,
\end{eqnarray}
where ${\bf F}_{1}({\bf r-r'})=\nabla_{r} f_{1}({\bf r-r'})$ and $\mbox{\boldmath $v$}({\bf r})$ is
the total velocity operator. These "statistical gauge potentials" tend to reduce the external
fields. As a result, for electronic filling factors $\nu = m/(pm\pm 1)$ and within the mean-field
approximation, the composite fermions filling factor in the residual magnetic field is
$\nu_{CF}=\pm m$ with a single occupied Landau level in each layer [the minus sign corresponds to
the fractions $\nu = m/(pm-1)$ for which the direction of the residual field is opposite to that of
the external field].

The resulting Lagrangian density reads\cite{Blok90,note1}
\begin{widetext}
\ba
\label{lagrangian1}
\nonumber {\cal L}&=&\sum_{I=1}^{m} \left\{ \psi_{I}^{*} \left[i\hbar\partial_{t}+e\left(A_{0}+a_{0}
\right)\right] \psi_{I} -\frac{1}{2M}\psi_{I}^{*}\left[ -i\hbar{\bf \nabla}+\frac{e}{c}
\left({\bf A}+{\bf a}\right)\right]^{2}\psi_{I}\right\} \\
&&-\frac{1}{2}\int d^2r' \rho({\bf r})U({\bf r-r'}) \rho({\bf r'}) -\frac{e}{2p\phi_{0}}\epsilon
^{\mu\nu\sigma}a_{\mu}\partial_{\nu}a_{\sigma} \; ,
\ea
\end{widetext}
where $M$ is the band mass. Here we have assumed that there is no tunneling between different layers.
We will comment on the effects of such tunneling processes when we discuss the electronic propogator
at the edge.

For a single-layer $\nu=m/(pm \pm 1)$ FQHE system the above transformation results, on the
mean-field level, in $\pm m$ filled composite fermions Landau levels within the layer. One can use
the Landau wave-functions in the residual field (and in the presence of the edge) to expand the
composite fermion operator $\psi({\bf r})=\sum_{I,k} c_{I,k} L^{-1/2}e^{iky} u_{I}(x-kl'^2) \equiv
\sum_{I}\psi_{I}({\bf r})$. The operators $\psi_I$ defined here are not fermionic. Although
operators which correspond to different Landau levels anti-commute, operators on the same level
obey $\{\psi_{I}({\bf r}),\psi_{J}^{\dagger}({\bf
r'})\}=\delta_{IJ}\Delta_I(x,x',y-y')\neq\delta_{IJ}\delta({\bf r}-{\bf r'})$. For example, away
from the edge $\Delta_1(x,x',y-y')=1/(2\pi l'^2)\exp[-(1/4)({\bf r}-{\bf
r'})^2+(i/2)(x+x')(y-y')]$, where ${\bf r}$ is measured in units of the magnetic length in the
residual field $l'=\sqrt{m/\nu}\,l$. Since the extent of $\Delta_I$ in the bulk is roughly
$\sqrt{I}l'$ and as long as the length scales under consideration are larger than $l'$ and $I$ is
not too large, one may try to approximate the anti-commutator (after an appropriate rescaling of
the fields) by a delta-function. It should be noted, however, that the integral over ${\bf r}$ of
$\Delta_I$ vanishes for even $I$ and it is not clear to what extent such an approximation is valid
(see also in this context Ref. \onlinecite{Zulicke03}).
Ignoring this issue one is able to recover the Lagrangian density (\ref{lagrangian1}) provided that
off-diagonal terms in the index $I$, corresponding to transitions between different Landau levels,
are neglected. Since these terms are not smaller than the diagonal ones such a step is not a priori
justified.

\begin{figure}[h!!!]
\setlength{\unitlength}{1in}
\psfig{figure=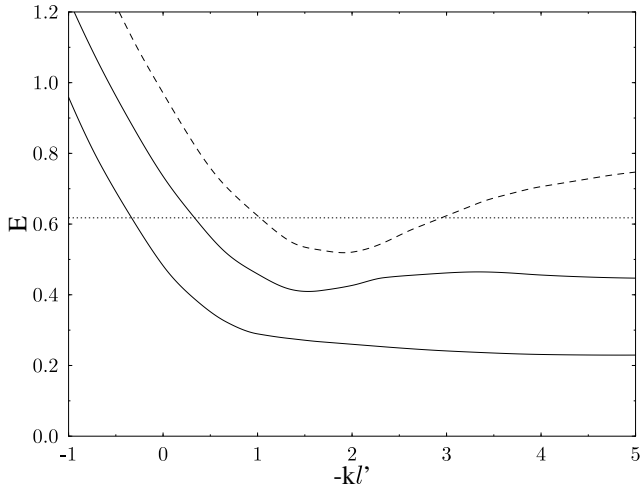,angle=270,width=3.34in}
\caption{A schematic plot of the composite fermions Landau levels for a sharp
$\nu=2/3$ edge. The dotted line represents the Fermi energy and $k$ denotes the conserved $y$
momentum of the states. When the edge is made smoother higher Landau levels (indicated by the
dashed line) may cross the Fermi energy at the vicinity of the edge.}
\label{fig-levels}
\end{figure}

In order to simplify the analysis we will assume that the confining potential is sharp enough so
that the filled Landau levels extend from the edge all the way into the bulk (see Fig.
\ref{fig-levels}). Cleaved edge overgrown samples appear to be good realizations of this
limit\cite{Huber05}. As the confining potential is made smoother higher Landau levels may cross the
Fermi energy in the vicinity of the edge \cite{Brey94,Chklovskii95} or the edge may reconstruct by
creating a strip of electrons separated from the bulk of the sample \cite{Chamon94,Wan02}. Under such
conditions one expects to find extra edge channels and edge modes. These cases will not be
addressed in the present work and we will consider $m$ electronic liquids (one per layer) that
occupy the entire plane of the system.

Next, we transform each of the fermionic fields $\psi_{I}$ into a bosonic field $\phi_{I}$
through an attachment of a single flux line in the $\pm\hat{\bf{z}}$ direction, namely,
\be
\label{cstransform2} \phi_{I}({\bf r})=e^{i\Lambda_{I}({\bf r})}\psi_{I}({\bf r}) \;,\;\;\;\;
\Lambda_{I}({\bf r})=\int d^2r' f_{2}({\bf r-r'})\rho_{I}({\bf r'}) \; ,
\ee
with $f_{2}({\bf r-r'})=f_{2}({\bf r'-r})\pm\pi$. The new gauge potentials introduced by this
transformation
\ba
\label{axy2}
&& {\bf a}_{I}({\bf r})=\mp\frac{\hbar c}{e}\int d^2r' {\bf F}_{2}({\bf r-r'})\rho_{I}({\bf r'})
\; , \\
\label{a02} && a_{I_{0}}({\bf r})=\mp\frac{\hbar}{e}\int d^2r' {\bf F}_{2}({\bf r-r'})\rho_{I}
({\bf r'})\mbox {\boldmath $v$}_{I}({\bf r'}) \; ,
\ea
where ${\bf F}_{2}({\bf r-r'})=\nabla_{r} f_{2}({\bf r-r'})$, cancel on the average the residual
magnetic field in the bulk of the sample. Their dynamics is controlled by an additional CS term
$\mp (e/2\phi_{0})\sum_{I=1}^{m}\epsilon^{\mu\nu\sigma}a_{I_{\mu}}\partial_{\nu}a_{I_{\sigma}}$
in the Lagrangian density (\ref{lagrangian1}). Note that contrary to the flux lines attached by
the first transformation (\ref{cstransform1}) the new ones are felt only by particles in the same
condensate as the particle carrying the flux.

By shifting $a_{I_{\mu}}$ to $a_{I_{\mu}}+a_{\mu}$, integrating over $a_{\mu}$ and introducing
the phase and velocity fields
\be
\label{vdef}
\phi_{I}=\sqrt{\rho_{I}}e^{i \theta_{I}} \; , \;\;\;
\mbox{\boldmath $v$}_{I}=\frac{\hbar}{M}{\bf \nabla}\theta_{I}+\frac{e}{Mc}
({\bf a}_{I}+{\bf A}) \; ,
\ee
we obtain the final hydrodynamic form of the Lagrangian and Hamiltonian densities \cite{Morinari96}
\begin{widetext}
\ba
\label{lagrangian}
\nonumber {\cal L}= &&\sum_{I=1}^{m} \left[ \rho_{I}\left(A_0+a_{I_{0}}-\partial_t\theta_{I}\right)
-\frac{1}{2}\rho_{I}\mbox{\boldmath $v$}_{I}^{2} -
\frac{1}{8}\frac{\left({\bf \nabla}\rho_{I}\right)^{2}}{\rho_{I}} \right]
-\frac{1}{{4\pi}}\sum_{I,J=1}^{m}\lambda_{IJ}\rho_{I}\rho_{J}
-\frac{1}{{4\pi}}\int d^2r' \rho({\bf r})U({\bf r-r'}) \rho({\bf r'}) \\
&&+\sum_{I,J=1}^{m}\left(K^{-1}\right)_{IJ}\left[\frac{1}{2}\epsilon^{ij}\partial_{t}v_{I}^{i}
\left(2\partial_{j}\theta_{J}-v_{J}^{j}\right)
+a_{I_{0}}\left(\epsilon^{ij}\partial_{i}v_{J}^{j}-1\right)\right] \; ,
\ea
\ba
\label{hamiltonian}
{\cal H}= &&\sum_{I=1}^{m} \left[\frac{1}{2}\rho_{I}\mbox{\boldmath $v$}_{I}^{2}
+\frac{1}{8}\frac{\left({\bf \nabla}\rho_{I}\right)^{2}}{\rho_{I}}-\rho_{I}A_{0} \right]
+\frac{1}{{4\pi}}\sum_{I,J=1}^{m}\lambda_{IJ}\rho_{I}\rho_{J}
+\frac{1}{{4\pi}}\int d^2r' \rho({\bf r})U({\bf r-r'}) \rho({\bf r'}) \; .
\ea
\end{widetext}
\noindent Henceforth length is measured in units of the magnetic length $l$, time is normalized by
the inverse cyclotron frequency $\omega_{c}=eB/Mc$, energy by $\hbar\omega_{c}$ and the density by
the Landau level degeneracy $\rho_{0}=1/2\pi l^{2}$. The above densities have been derived under
the assumption that no localized vortices are present in the sample. Anticipating the use of the
mean-field approximation we have included in (\ref{lagrangian},\ref{hamiltonian}) additional
contact interactions in order to partially account for the kinetic energy coming from the short
range structure of the wave-function which is absent in this approximation \cite{Ezawa92,Orgad96}.
We comment on the values of their couplings $\lambda_{IJ}$ below.
The matrix $\left(K^{-1}\right)_{IJ}=\mp p/(pm\pm 1) \pm \delta_{IJ}$ is the inverse of
$K=pC\pm {\bf 1}$ with ${\bf 1}$ and $C$ being the identity and pseudo-identity matrices ($C_{IJ}=1$).
The matrix $K$ is an example of the general classification scheme, introduced by
Wen and Zee \cite{Wen92,Wen95}, of Abelian FQHE states. Its elements $K_{IJ}$ are
the number of flux lines affecting the phase of a particle in the
$I$-th condensate when it encircles a particle in the $J$-th condensate.

\section{Mean-Field Solutions}
\label{sec:MFAsol}

Variation of the action (\ref{lagrangian}) with respect to the variables $a_{I_{0}},
\mbox{\boldmath $v$}_{I},\theta_{I}$ and $\rho_{I}$ results in the mean-field equations
\begin{subequations}
\ba
\label{bfield} &&\hspace{-1cm}
\sum_{J=1}^{m}\left(K^{-1}\right)_{IJ}\epsilon_{ij}\partial_i v_{J}^{j}=
\frac{\nu}{m}
-\rho_{I} \; ,\\
\label{efield} &&\hspace{-1cm}
\sum_{J=1}^{m}\left(K^{-1}\right)_{IJ}\left(\partial_t\partial_i\theta_{J}
-\partial_t v_{J}^{i}-\partial_i a_{J_{0}}\right)=\epsilon_{ij}\rho_{I} v_{I}^j \; , \\
\label{cont}
&&\hspace{-1cm}
\partial_t\rho_{I}=-{\bf \nabla}(\rho_{I} \mbox{\boldmath $v$}_{I}) \; ,\\
\label{sch} \nonumber &&\hspace{-1cm}
\partial_t\theta_{I}=-\frac{1}{2}\mbox{\boldmath $v$}_{I}^{2}+\frac{1}{2}\frac{{\bf \nabla}^2
\sqrt{\rho_{I}}}{\sqrt{\rho_{I}}}+A_0+a_{I_{0}}\\
&&\hspace{0.1cm} -\frac{1}{2\pi}\sum_J\lambda_{IJ}\rho_J
-\frac{1}{{2\pi}}\int U({\bf r-r'})\rho({\bf r}')d^2 r' .
\ea
\end{subequations}

In the absence of confining potentials these equations support a uniform solution $\rho_{I}=\nu/m$ with
$\mbox {\boldmath $v$}_{I}=a_{I_0}=0$ and $\theta_I=-\mu t$. The correct noninteracting ($U=0$)
energy density ${\cal H}=\nu\hbar\omega_c/2$ for this solution is recovered by taking
$\lambda_{IJ}=2\pi m/\nu \delta_{IJ}$. One also finds $\mu_{I}=\hbar\omega_c$
indicating that mean-field configurations with slightly different filling factor and which are
continuously connected to the constant solution necessarily involve higher Landau levels\cite{note2}.
This choice is not unique. Taking $\lambda_{IJ}=2\pi K_{IJ}$ also
recovers the correct noninteracting energetics. As we shall see, the form of $\lambda_{IJ}$ affects some
of the details of the mean field solutions. It does not, however, modify any of the general conclusions
of our study.

Up to this point we have not specified the kernels $f_{1,2}$ of the CS transformations
(\ref{cstransform1},\ref{cstransform2}) or equivalently fixed the gauge for the statistical
potentials $a_{\mu}$ and $a_{I_{\mu}}$. The canonical choice
\be
\label{f} f_{1,2}({\bf r-r'})=\alpha_{1,2}\arctan\left(\frac{y-y'}{x-x'}\right) \; ,
\ee
with $\alpha_{1,2}=p,\pm 1$, induces, through ${\bf F}_{1,2}({\bf r-r'})=\lim_{\varepsilon
\rightarrow 0^{+}}\,\alpha_{1,2}\,\hat{\bf z}\times ({\bf r-r'})/(|{\bf r-r'}|^2+\varepsilon)$, a
symmetric gauge which is in conflict with the translational symmetry of the problem in the $y$
direction. We are free, however, to add to $f_{1,2}$ functions $g_{1,2}$ satisfying $g_{1,2}({\bf
r-r'})=g_{1,2}({\bf r'-r})$, corresponding to a regular gauge transformation in the Lagrangian
formalism. We choose
\ba
\label{g} \nonumber g_{1,2}({\bf r-r'})&=&\lim_{\varepsilon\rightarrow
0^{+}}\alpha_{1,2}\frac{y-y'}{\sqrt{(y-y')^2+
\varepsilon}}\\
&\times& \arctan\left[\frac{x-x'}{\sqrt{(y-y')^2+\varepsilon}}\right] \; .
\ea
As a result $(f+g)_{1,2}({\bf r-r'})=\pi\alpha_{1,2}[1-\epsilon(y-y')]$ with $\epsilon(x)=\Theta(x)-1/2$
where $\Theta(x)$ is the step function. $(f+g)_{1,2}$ has a branch cut which we
take to be along the positive $x-x'$ axis and therefore $\nabla_r (f+g)_{1,2}=[0,2\pi\alpha_
{1,2}\epsilon(x-x')\delta(y-y')]$. Accordingly, the statistical potentials [after the shift
performed in obtaining (\ref{lagrangian})] are in the Landau gauge
\ba
\label{landauaxy}  {\bf a}_{I}({\bf r})&=&\left[0\, , -\int dx' \sum_{J=1}^{m} K_{IJ}
\epsilon(x-x')\rho_{J}(x',y)\right] , \\
\label{landaua0} \nonumber  a_{I_{0}}({\bf r})&=&-\int dx' \sum_{J=1}^{m} K_{IJ}\epsilon(x-x')
\rho_{J}(x',y) v_{J}^{y}(x',y) \; . \\
\ea

\begin{figure}[ht!!!]
\setlength{\unitlength}{1in}
\psfig{figure=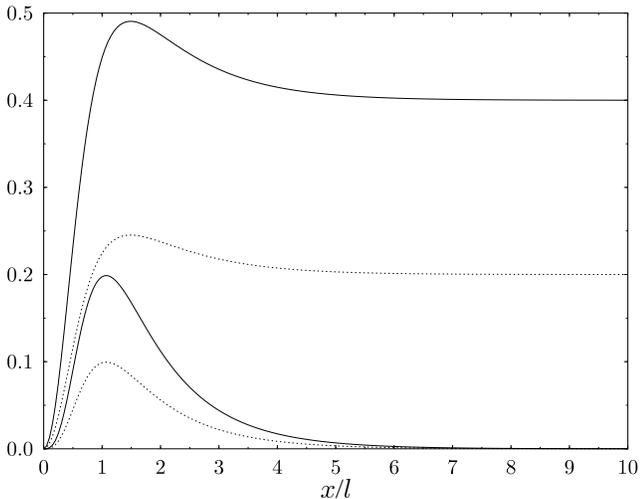,angle=270,width=3.3in}
\caption{Mean-field solution for a $\nu=2/5$ edge in equilibrium
$\kappa_1=\kappa_2=0$ with short range interaction $\lambda_s=1$ and $\lambda_{IJ}=2\pi m/\nu\delta_{IJ}$.
The upper dotted and solid lines depict the densities $\rho_1=\rho_2$ and $\rho=\rho_1+\rho_2$ respectively,
in units of $\rho_0$. The lower lines correspond to the current densities $j^y_1=j^y_2$ and
$j^y=j^y_1+j^y_2$ in units of $\rho_0\omega_c l$.} \label{fig-sr2/5}
\end{figure}

\begin{figure}[ht!!!]
\setlength{\unitlength}{1in}
\psfig{figure=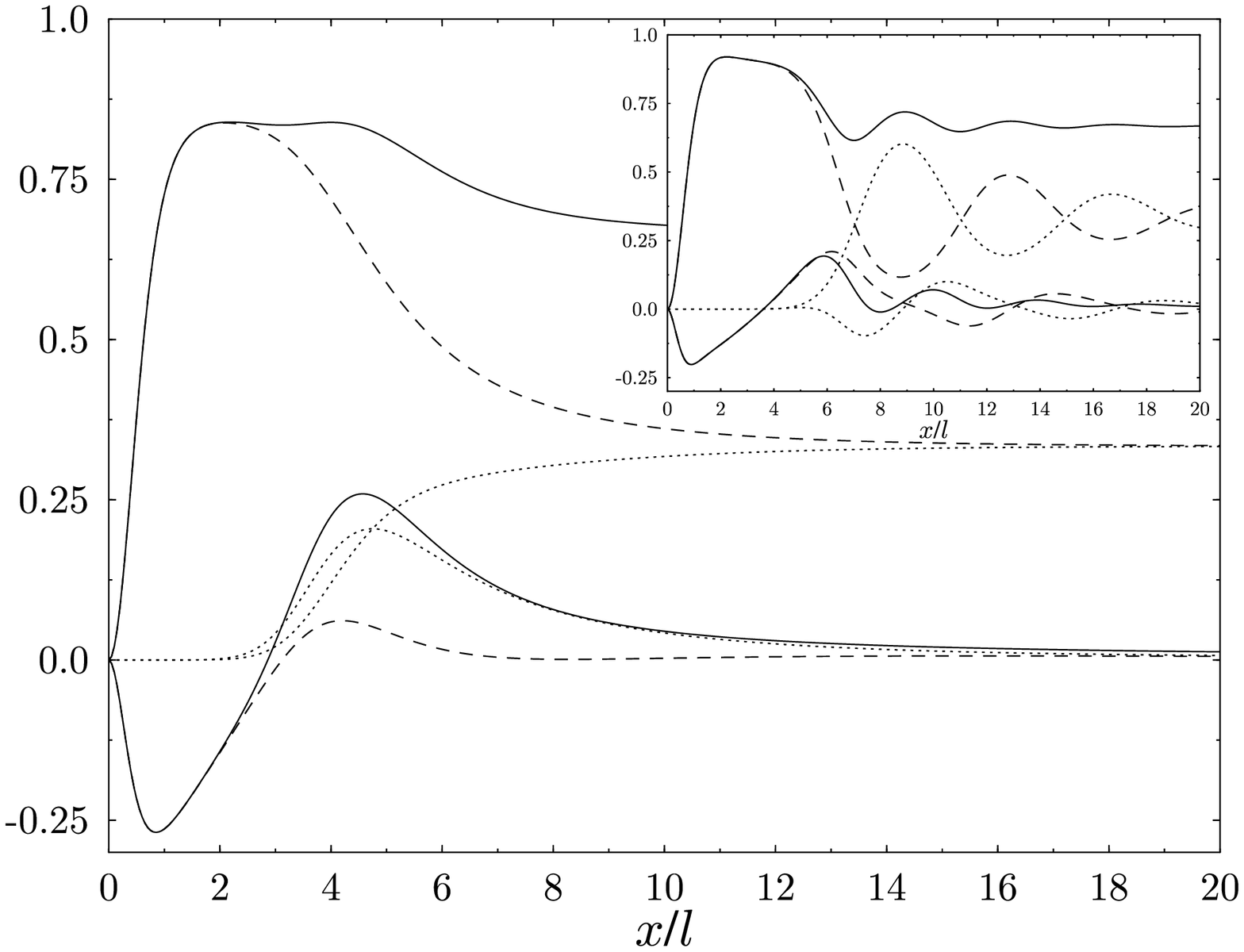,angle=270,width=3.3in}
\caption{Mean-field solution for a $\nu=2/3$ edge out of equilibrium $\kappa_1=3$,
$\kappa_2=-1$ with Coulomb interaction \protect{\cite{note3}} $\lambda_c=0.2$ and
$\lambda_{IJ}=2\pi m/\nu\delta_{IJ}$. The upper dotted,
dashed and solid lines depict the densities $\rho_1$, $\rho_2$ and $\rho=\rho_1+\rho_2$
respectively, in units of $\rho_0$. The lower lines correspond to the current densities $j^y_1$,
$j^y_2$ and $j^y=j^y_1+j^y_2$ in units of $\rho_0\omega_c l$. For increased asymmetry between the two
condensates the strip of enhanced density near the edge becomes wider with filling
factor approaching 1, cf. Ref. \protect{\onlinecite{MacDonald90}}. The solution corresponding to
the same parameters but when $\lambda_{IJ}=2\pi K_{IJ}$ is shown in the inset.} \label{fig-c2/3}
\end{figure}

We will find it more convenient to obtain the static solutions and the edge modes in a slightly
different gauge where we replace $\epsilon(x-x')$ in (\ref{landauaxy},\ref{landaua0}) by
$\Theta(x-x')$. Using this gauge we look for static solutions of the form
\be
\label{sol}
\theta_{I}=\kappa_{I}y-\mu_{I}(\{\kappa_{I}\})t \;,\;\;\;\;\; \rho_{I} = \rho_{I}(x) \; .
\ee
Taking for definiteness the confining potential to be an infinite wall situated at $x\leq 0$ we
set the density to zero on the edge. For the external vector potential we use the gauge ${\bf
A}=[0,Bx]$ which gives velocity fields of the form $v_{I}^{x}=0$ and $v_{I}^{y}=v_{I}^{y}(x)$.
Assuming that $a_{I_{0}}$ are also functions of $x$ only and inserting the above ansatz into
(\ref{bfield})-(\ref{sch}) we obtain a set of coupled equations for the various fields appearing in
the problem. We solve the equations numerically for the cases of short-range interaction $U=2\pi
\lambda_s\delta({\bf r-r'})$ and Coulomb interaction $U=(2\pi\lambda_c)/|{\bf r-r'}|$ with
$\lambda_c=e^2/(2\pi\epsilon l\hbar\omega_c)$ plus a  constat positive neutralizing background.
Representative solutions are presented in Figs. \ref{fig-sr2/5} and \ref{fig-c2/3}

The solutions to the mean-field equations form an $m$-dimensional manifold which is parameterized
by the quantum numbers  $\{\kappa_{I}\}$. These quantum numbers are the conserved momenta along
the edge of the different condensates and may be thought of as their collective guiding center
coordinates. Varying their values translates the condensates in the $x$ direction. They also
determine the amount of excess charge (per unit length in the $y$ direction) carried by each of the
condensates relative to a step-like constant density profile.  To demonstrate this relation we
integrate Eq. (\ref{bfield}), noting that the ansatz (\ref{sol}) implies $v_{I}^{y}(0)= \kappa_{I}$
and that the velocity vanishes in the bulk of the system. Consequently we obtain
\be
\label{kmeaning}
\sum_{J=1}^{m}\left(K^{-1}\right)_{IJ}\kappa_{J}=\int\left(\rho_{I}-\frac{\nu}{m}\right)dx \; .
\ee

For a given set $\{\kappa_{I}\}$ the values of $\{\mu_{I}\}$ are determined by the requirement
that far from the edge the density attains its bulk value $\nu$. To appreciate their physical
meaning we consider the energy difference between two neighboring mean-field solutions differing
by the number of particles in one of the condensates. For our ansatz such a variation is achieved
by a suitable change of $\{\kappa_{I}\}$ using (\ref{kmeaning}). Utilizing the mean-field
equations one finds in this case
\ba
\label{evariation}
\nonumber
\delta E &=& \int d^2 r \sum_{I=1}^{m} \left( \frac{\delta {\cal H}}{\delta\rho_{I}}
\delta\rho_{I} + \frac{\delta {\cal H}}{\delta \mbox{\boldmath $v$}_{I}}
\delta\mbox{\boldmath $v$}_{I} \right) \\
&=&-\int d^2 r \sum_{I=1}^{m} \partial_{t}\theta_{I} \delta\rho_{I} =
\sum_{I=1}^{m} \mu_{I}\delta N_{I} \; ,
\ea
where $\int d^2 r \delta\rho_{I} \equiv \delta N_{I}$.
Accordingly, $\mu_{I}$ is interpreted as the chemical potential of the $I$-th condensate.
In the noninteracting limit we find numerically that the
chemical potential of equilibrium configurations (see below) decreases and approaches $\hbar\omega_c/2$
as the edge is placed away from the wall. This reflects the fact that under
such conditions neighboring ground states may be related by the
addition of charge to the rim of the electronic droplet. This result thus provides further evidence
for the utility of the added contact interactions in the mean-field Hamiltonian.

In case of short-range interaction the fields approach their bulk values exponentially in the
distance from the wall \cite{Orgad96} (cf. Fig. \ref{fig-sr2/5}). As a result we find from Eq.
(\ref{sch}) that for large $x$ $a_{I_{0}}=-\mu_{I}+\nu\lambda_s+1$. Noting that in our gauge
$a_{I_{0}}(0)=0$ we integrate Eq. (\ref{efield}) to obtain the following relation connecting the
chemical potentials $\{\mu_{I}\}$ to the currents carried by the different condensates
\be
\label{srmu}
\sum_{J=1}^{m}\left(K^{-1}\right)_{IJ}\mu_{J}-\frac{\nu}{m}(1+\nu\lambda_s)=I_{I} \; .
\ee

For Coulomb interaction the asymptotic behavior of the fields
at large distances from the edge is determined by the total charge carried by the system relative
to the constant neutralizing background. From (\ref{kmeaning}) we find the linear charge density
along the edge to be $\sigma=(\nu/m)\sum_{I=1}^{m}\kappa_{I}$. Assuming that the length of the sample
$L$ is much larger than the length over which the density changes appreciably ($\sim \sigma/\nu$) we
can integrate the interaction term in (\ref{sch}) to find $(1/2\pi)\int U({\bf r-r'})
[\rho({\bf r}')-\nu]d^2 r'\simeq 2\lambda_c \sigma \ln L -2\lambda_c\int[\rho(x')-\nu]\ln(|x-x'|)dx'$.
The constant term is the leading contribution to the chemical potentials. It corresponds to the
dominant part of the charging energy of the edge $\lambda_c \sigma^2 L\ln L$. The remaining contributions
to the chemical potentials will be denoted by $\tilde\mu_{I}=\mu_{I}-2\lambda_c \sigma \ln L$. When
$\sigma\neq 0$ we find at large distances $\rho_{I}\simeq \nu/m+2\lambda_c \sigma/x^{2}$,
$v_{I}^{y}\simeq 2\lambda_c \sigma/x$ and $a_{I_{0}}\simeq -\tilde\mu_{I}+1-2\lambda_c \sigma\ln x$
(with or without relative oscillations between the condensates, cf. Fig \ref{fig-c2/3}). The slow decay
of the current density results in a logarithmically diverging
contribution to the integrated current from the wall to a point $x$. Defining
$\tilde I_{I}=I_{I}-2\lambda_c \sigma(\nu/m)\ln x$ one obtains
\begin{equation}
\label{comu}
\sum_{J=1}^{m}\left(K^{-1}\right)_{IJ}\tilde\mu_{J}-\frac{\nu}{m}=\tilde I_{I} \; .
\end{equation}
If $\sigma=0$ the fields relax to their bulk values more rapidly $\rho_{I}\sim \nu/m+ x^{-3}$,
$v_{I}^{y}\sim x^{-2}$ and $a_{I_{0}}\sim -\mu_{I}+1+x^{-1}$. Consequently the current is finite and
a similar relation to (\ref{comu}) holds for $\{\mu_{I}\}$ and $\{I_{I}\}$.

We remark that (\ref{comu}) leads to the appropriate quantized Hall conductance for a wide bar
\cite{Orgad96} ($L\gg W\gg\kappa l$). Consider two edges in equilibrium with chemical potentials
$\mu_{1,2}$ and charge densities $\pm \sigma$. Summing over the edge channels we find
\begin{equation}
\label{hallquant}
I=I_1-I_2=\nu[\mu_1-\mu_2+4\lambda_c \sigma \ln (W/L)]\equiv \nu V_{\rm Hall} \; .
\end{equation}
The expression for the Hall voltage reflects the modification of the edge chemical potential due to
the electrostatic potential induced by the opposite edge.

\begin{figure}[ht!!!]
\setlength{\unitlength}{1in}
\psfig{figure=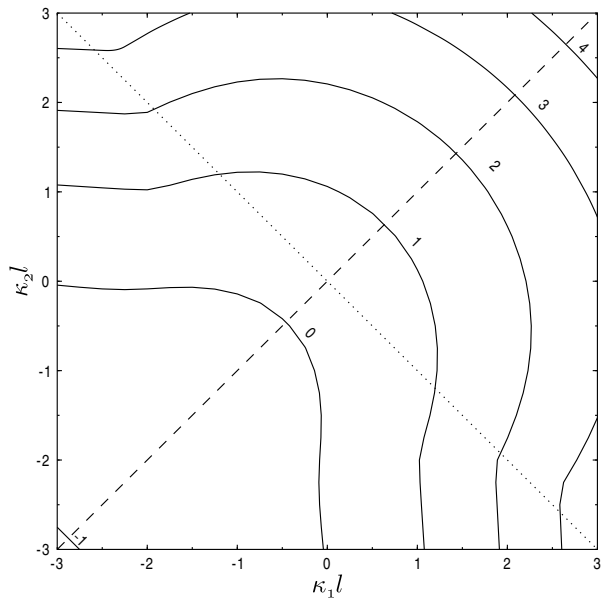,angle=270,width=3.2in}
\vspace{-0.2cm}
\caption{A map of the mean-field solutions for a $\nu=2/5$ system with Coulomb
interaction $\lambda_c=0.2$ and $\lambda_{IJ}=2\pi m/\nu\delta_{IJ}$. The solid curves correspond
to equi-current families of solutions. The integrated
current over a distance of $80l$ from the edge is given in units of
$e\omega_c$. The dashed and dotted lines are the equilibrium $(\mu_1=\mu_2)$ and neutrality
$(\kappa_1=-\kappa_2)$ lines respectively. The reversal in the direction of the current with
decreasing $\kappa$ signals a transition from skipping orbits along the wall to circular orbits on
the rim of the Hall droplet.} \label{fig-map2/5}
\end{figure}

\begin{figure}[ht!!!]
\setlength{\unitlength}{1in}
\vspace{-0.1cm}
\psfig{figure=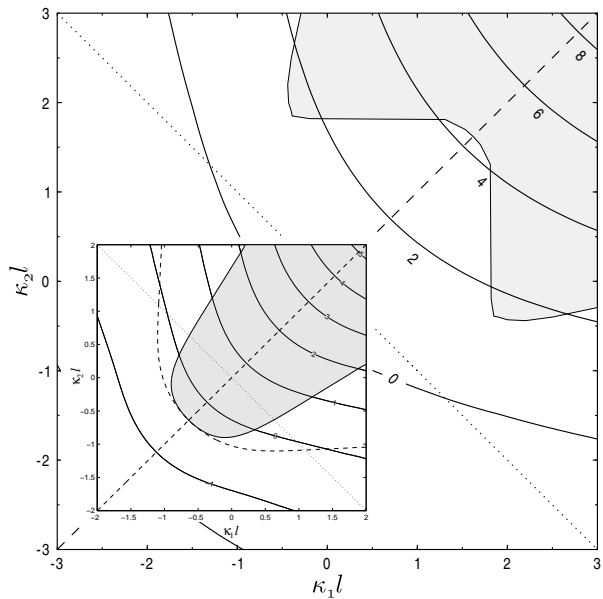,angle=270,width=3.2in}
\vspace{-0.2cm}
\caption{A map of the mean-field solutions for a $\nu=2/3$ system with Coulomb
interaction $\lambda_c=0.2$ and $\lambda_{IJ}=2\pi m/\nu\delta_{IJ}$. The notations are similar to
Fig. \ref{fig-map2/5}. In the shaded region the mean-field solutions are unstable, as described in
the text. The inset shows a similar map for $\lambda_{IJ}=2\pi K_{IJ}$. Note the
bifurcation of the equilibrium line $\kappa_1=\kappa_2$ upon entering the
unstable region and the new asymmetric equilibrium states.}
\label{fig-map2/3}
\end{figure}

The particular solution to be realized out of the manifold of mean-field configurations is
determined by the external constraints imposed on the system. Strictly speaking, since the
$\kappa_I$ are conserved quantities, the entire manifold corresponds to equilibrium states
of the system. Taking, however, a more experimentally motivated point of view we will identify
the equilibrium configurations as those which minimize the energy for a given total charge
at the edge, as determined by the electrostatic potentials applied to the edge.
Any perturbation which breaks the translation invariance in the $y$ direction will cause
the system to seek and eventually settle down in this solution.
A similar definition may be applied to the case where the edge is constrained to sustain a
given total current instead of charge.

As can be seen from Eq.
(\ref{evariation}) the symmetrical configurations $(\kappa_I=\kappa, \mu_I=\mu)$ are extrema of the
energy functional for a fixed total charge at the edge $(\sum_I\delta N_I=0)$. From our numerical
results we find that as long as this charge is not too large they indeed correspond to
equilibrium solutions, see Figs. \ref{fig-map2/5} and \ref{fig-map2/3}. However, when the charge
is increased the symmetrical solutions turn into local energy maxima and new asymmetrical
equilibrium states appear. This is particularly clear in the case of the $\nu=2/3$ edge, see Fig.
\ref{fig-map2/3}, but the effect also exists for large $\kappa$ in the $\nu=2/5$ edge.

In the following section we will obtain the gapless edge excitations. As we have mentioned in the
introduction they consist of a single magnetoplasmon mode and $m-1$ acoustic branches. For a
$\nu=2/3$ edge we find that the latter become soft inside a parameter region, denoted by the shaded
area in Fig. \ref{fig-map2/3}, whose extent depends on the particular model interactions
(see also the inset in Fig. \ref{fig-exc2/5}).
Since in the acoustic mode the two condensates oscillate out of phase this softening signals a
tendency of the edge to reconstruct by transferring charge from one condensate to the other. i.e.
by deviating from the symmetrical solution along the direction of the neutrality line. Fig.
\ref{fig-map2/3} confirms this conclusion. We note that in reality the edge reconstruction
may involve population of additional Landau levels in the vicinity of the edge. Such
processes, however, fall outside of the model studied here.

When no constraints are imposed on the system it will choose the globally lowest available state. In the
case of Coulomb interaction such a solution corresponds to a neutral configuration ($\kappa_I=0$)
that minimizes the dominant charging energy (in realistic edges, for example in overgrown cleaved
edge samples, additional potentials due to near-by charges may alter this
conclusion\cite{Levitov01}). For short-range interaction the chemical potential of a system in
equilibrium is a monotonically increasing function of $\kappa$. Consequently energy is gained by
placing the edge as far as possible from the confining wall. In order to fix the edge one has to
introduce an additional confining potential.

The equilibrium solutions compose only a small fraction of the mean-field manifold. The rest of the
solutions describe possible realizations of the edge out of equilibrium. Deliberate action must be
taken in order to place the system in one of these configurations. Experimentally, this may be
achieved through selective population of edge channels employing gates over the edge. The
relatively large equilibration lengths found in experiments \cite{Kouwenhoven90,Chang92} (few $\mu
{\rm m}$) raise the possibility of observing such states.

\section{Symmetrical Edges}
\label{sec:Sym}

\subsection{Edge Excitations}
\label{subsec:Sym-excite}

Having obtained the mean-field solutions we are now in a position to study small fluctuations about
them. To this end we consider the RPA equations which are derived from Eqs.
(\ref{bfield})-(\ref{sch}) by linearization around one of the mean-field configurations

\begin{figure}[ht!!!]
\vspace{0.2cm}
\setlength{\unitlength}{1in}
\psfig{figure=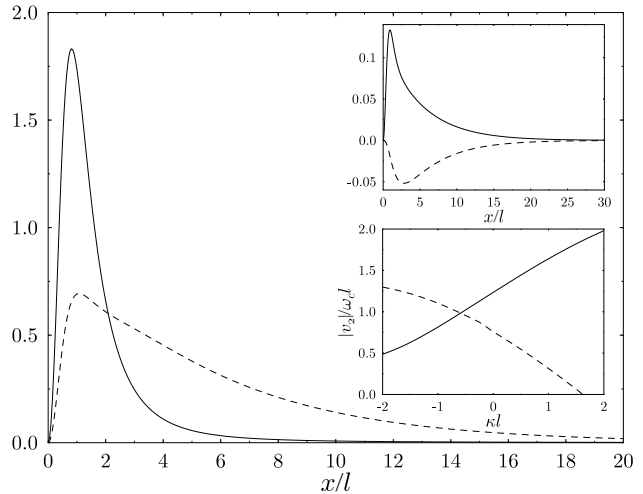,angle=270,width=3.3in}
\vspace{0.2cm}
\caption{Excitations of a $\nu=2/5$ edge in equilibrium $\kappa_1=\kappa_2=0$
with Coulomb interaction $\lambda_c=0.2$ and $\lambda_{IJ}=2\pi m/\nu\delta_{IJ}$. The solid line
depicts the density profile of the magnetoplasmon mode where $\delta\rho_1^1=\delta\rho_2^1$.
The dashed line corresponds to the density profile of the neutral
mode $\delta\rho_1^2=-\delta\rho_2^2$ where the condensates oscillate out of phase.
The upper inset presents the translation modes of this edge.
The solid and dashed curves correspond to $\partial\rho_1/\partial\kappa_1=\partial\rho_2/\partial
\kappa_2$ and $\partial\rho_1/\partial\kappa_2=\partial\rho_2/\partial\kappa_1$ respectively.
The lower inset shows the velocity of the neutral mode for symmetric edges
$\kappa_1=\kappa_2=\kappa$ with the same interaction. The solid and dashed lines correspond to
a $\nu=2/5$ and $\nu=2/3$ edges respectively.}
\label{fig-exc2/5}
\end{figure}

\begin{subequations}
\ba
\label{lbfield}
&&\hspace{-1.0cm}\sum_{J=1}^{m}\left(K^{-1}\right)_{IJ}\epsilon_{ij}\partial_i \delta v_{J}^{j}=
-\delta\rho_{I} \; ,\\
\label{lefield} \nonumber
&&\hspace{-1.0cm}\sum_{J=1}^{m}\left(K^{-1}\right)_{IJ}\left(\partial_t\partial_i\delta\theta_{J}
-\partial_t \delta v_{J}^{i}-\partial_i \delta a_{J_{0}}\right)= \\
&&\hspace{2.48cm}\epsilon_{iy} \delta\rho_{I} v_{I}^y + \epsilon_{ij}\rho_{I}
\delta v_{I}^j  \; ,\\
\nonumber \\
\label{lcont}
&&\hspace{-1.0cm}\partial_t\delta\rho_{I}=-{\bf \nabla}(\delta\rho_{I} \mbox{\boldmath $v$}_{I} +
\rho_{I} \mbox{\boldmath $\delta v$}_{I}) \; ,\\
\label{lsch} \nonumber
&&\hspace{-0.99cm}\partial_t\delta\theta_{I}=-v_{I}^{y}\delta
v_{I}^{y}+P(\rho{_I},\delta\rho{_I})
+\frac{\partial_{y}^2\delta\rho_{I}}{4\rho_{I}}+\delta a_{I_{0}}\\
\nonumber
&&\hspace{0.3cm}-\frac{1}{2\pi}\sum_J\lambda_{IJ}\delta\rho_J
-\frac{1}{2\pi}\int U({\bf r-r'})\delta\rho({\bf r}')d^2 r'\; . \\
\ea
\end{subequations}
Here $P(\rho{_I},\delta\rho{_I})$ is the part of the linearized "quantum pressure" term, i.e., the
second term on the right-hand side of (\ref{sch}) containing $x$ derivatives, and
$\delta\rho=\sum_{I=1}^{m}\delta\rho_{I}$.

We can easily identify $m$ independent solutions to the above equations. These are the small
variations that connect the static solution, around which we linearized, and its neighboring
configurations on the mean-field manifold. They are obtained from the static solution by taking its
derivatives with respect to the $m$ conserved momenta $\kappa_{I}$. Representative examples of
these translation modes are shown in the inset of Fig. \ref{fig-exc2/5}. Evidently they correspond
to a shift of the condensates in the $x$ direction and hence to the addition and removal of charge
from the vicinity of the edge. At present the amount of this charge is arbitrary since, due to the
linearity of (\ref{lbfield})-(\ref{lsch}), it is determined only up to a multiplicative constant.
However, we will show that the periodic boundary conditions impose on it discrete values in units
of $\nu/m$. We therefore interpret these modes as edge states of the $m$ types of
quasiparticles\cite{note4} that can be generated by threading one of the condensates with an extra
flux line [of the type associated with the CS transformation (\ref{cstransform2})].

The translation modes are the building blocks for the construction of the $m$ gapless excitation
branches supported by the edge of the system. Considering the long-wavelength limit we find
\ba
\label{lsol} \nonumber \delta\rho_I^{\alpha,k}&=&\sum_{J=1}^m A_J^\alpha(k)\frac{\partial\rho_I(x)}
          {\partial\kappa_J}e^{-i(ky-\omega_\alpha t)} \; ,\\
\nonumber \delta v_I^{x,\alpha,k}&=&\frac{i}{\rho_I(x)}\sum_{J,L=1}^m
A_J^\alpha(k)\left(K^{-1}\right)
           _{IL}\Biggl[\omega_\alpha\frac{\partial v_L^y(x)}{\partial\kappa_J} \\
\nonumber &&-k\frac{\partial a_{L_{0}}(x)}{\partial\kappa_J}-\omega_\alpha\delta_{LJ}
             \Biggr]e^{-i(ky-\omega_\alpha t)} \; ,\\
          \delta v_I^{y,\alpha,k}&=&\sum_{J=1}^m A_J^\alpha(k)\frac{\partial v_I^y(x)}
          {\partial\kappa_J}e^{-i(ky-\omega_\alpha t)} \; ,\\
\nonumber \delta a_{I_{0}}^{\alpha,k}&=&\sum_{J=1}^m A_J^\alpha(k)\frac{\partial a_{I_{0}}(x)}
             {\partial\kappa_J}e^{-i(ky-\omega_\alpha t)} \; ,\\
\nonumber \delta\theta_I^{\alpha,k}&=&\frac{i}{k}A_I^\alpha(k)e^{-i(ky-\omega_\alpha t)} \; ,
\ea
where $\alpha=1,\cdots,m$ is the branch index. The real and imaginary parts of (\ref{lsol}) are, to
first order in $k$, independent solutions of the RPA equations. The expansion coefficients
$A_J^\alpha(k)$ and the dispersion relations $\omega_\alpha(k)$ are determined from the secular
equation obtained by plugging (\ref{lsol}) into (\ref{lsch})
\begin{eqnarray}
\label{secular} \nonumber
\sum_{J=1}^{m}\left[B_{IJ}-\frac{\omega_\alpha}{k}\delta_{IJ}\right]A_J^\alpha(k)=0 \; , \\
B_{IJ}=\frac{\partial\tilde\mu_I}{\partial\kappa_J}+2\lambda_c\frac{\nu}{m}\ln\left
(\frac{2e^{-\gamma}}{|k|}\right) \; .
\end{eqnarray}
$\gamma$ is the Euler constant appearing in the long wavelength limit of the Coulomb interaction
term (This and the following results hold also in the case of short-range interaction after
replacing $\tilde\mu$ by $\mu$ and setting $\lambda_c=0$).

Due to the assumed symmetry of the mean-field solution
$\partial\tilde\mu_I/\partial\kappa_I\equiv\eta_1$ and $\partial\tilde\mu_I/\partial
\kappa_J\equiv\eta_2$ for all $J\neq I$. As a result the matrix $B$ is symmetric and takes the form
\begin{equation}
\label{beq} B=\left[\eta_2+2\lambda_c\frac{\nu}{m}\ln\left(\frac{2e^{-\gamma}}{|k|}\right)\right]C +
\left(\eta_1-\eta_2\right){\bf 1} \; .
\end{equation}
Here again $C$ is the pseudo-identity matrix.

The eigenvalue problem (\ref{secular}) is readily solved to give one edge-magnetoplasmon mode
and $m-1$ degenerate branches with acoustic dispersion\cite{Morinari96}. The magnetoplasmon
corresponds to a coherent in-phase oscillation of the condensates with $A_I^1(k)=2\pi/L\nu$
(cf. Fig. \ref{fig-exc2/5}). We have normalized the eigenvector such that the charge carried by
its density profile is $\int d^2r \sum_{I,J=1}^m A_J^1(k)(\partial\rho_I/\partial\kappa_J)= 2\pi$
(1 in dimensionful units) as can be easily verified using (\ref{kmeaning}). Its dispersion
relation is
\begin{equation}
\label{chargeq}
v_1\equiv\frac{\omega_1}{k}=\eta_1+(m-1)\eta_2+2\nu\lambda_c \ln\left(\frac{2e^{-\gamma}}{|k|}\right) \; .
\end{equation}
We will show below that this mode propagates in the positive $y$ direction for all fractions.

The rest of the modes are degenerate and have acoustic dispersion with velocity
\begin{equation}
\label{neutraleq}
v_2\equiv\frac{\omega_2}{k}=\eta_1-\eta_2 \; .
\end{equation}
They correspond to excitations in which the different condensates oscillate out of phase under
the condition $\sum_{I=1}^m A_I^{\alpha}(k)=0$ and are therefore neutral
(this perfect neutrality is not expected to hold in the case of a single-layer
system). In the following we take them to be orthogonal and normalized
according to $|{{\bf A}^\alpha}|=2\pi/L$.

The direction in which the neutral modes propagate depends on the
filling factor. The reason for this stems from the relation between
the velocity of the mode $v_2$ and the inverse compressibility of the edges of the
condensates. The latter can be viewed, according to Eq. (\ref{kmeaning}), as one-dimensional systems of
length $L$ and $N_I=L\sum_{J=1}^{m}\left(K^{-1}\right)_{IJ}\kappa_{J}$ particles (where negative values
of $N_I$ correspond to minus the number of holes on the edge). Consider
the inverse compressibility of the edge of the $I$-th condensate $\chi_I^{-1}=(N_I^2/L)(\delta
\mu_I/\delta N_I)$ under the constraint of fixed total charge, i.e. for variations satisfying
$\sum_J\delta\kappa_J=0$. Using Eq. (\ref{kmeaning}) and the form of $K^{-1}$ one
finds for the symmetric solutions $\chi^{-1}_I=\pm(N_I/L)^2(\eta_1-\eta_2)$. The stability requirement
$\chi_I>0$ then implies that ${\rm sgn}(v_2)=\pm 1$ for $\nu=m/(pm\pm 1)$.

We remark that for short range interaction the preceding analysis can be carried out to second
order in $k$. To this order the RPA solution is the same as (\ref{lsol}) but with $\delta\rho_I^
{\alpha,k}$, $\delta v_I^{y,\alpha,k}$ and $\delta a_{I_{0}}^{\alpha,k}$ multiplied by $(1+k)$.
As a result one finds for the dispersion relations
\ba
\label{srdisp}
\omega_1(k)&=&[\eta_1+(m-1)\eta_2]k(1+k) \; , \\
\omega_2(k)&=&(\eta_1-\eta_2)k(1+k) \; .
\ea
The inverse compressibility vanishes when the edge of the static solution is far
away from the wall (i.e at large negative $\kappa$) for which case $\omega_2\sim k^3$. A similar
result for the case $\nu=1$ was obtained in Ref. \onlinecite{Giovanazzi94}.

\subsection{The Tomonaga-Luttinger model}
\label{subsec:Sym-TL}

In addition to the gapless edge modes a general excited state of the system may include other types
of excitations such as bulk magnetoplasmons, magnetorotons, quasiparticles and quasiholes.
However, since all of them involve an energy gap
the edge modes are expected to dominate the low energy physics. We will thus
restrict our consideration to cases in which the deviations of the various fields from their static
ground state configurations can be expanded solely in terms of the gapless excitations. Our plan is to
plug these expansions into the Lagrangian (\ref{lagrangian}) in
order to obtain its low energy effective form and to use them to invert the CS transformations
(\ref{cstransform1},\ref{cstransform2}). The result will be an expression for the original electronic
operators in terms of the bosonic edge modes. For this purpose we are interested to use the kernels
(\ref{f},\ref{g}) and accordingly have to transform back to the gauge (\ref{landauaxy},
\ref{landaua0}) which they induce. As a result the gauge dependent quantities $a_{I_0}$ and
$\theta_I$ change, the latter to $[\kappa_I-1/2\int dx \sum_{J=1}^m K_{IJ}\rho_J(x)]y -
[\mu_I-1/2\int dx \sum_{J=1}^m K_{IJ}\rho_J v_J^y(x)]t$. The expansion of the deviation fields
is then
\ba
\label{expansion}
\nonumber
\delta\rho_I&=&\sum_{\alpha=1}^m\sum_k\rho_{\alpha}(k)\delta\rho_I^{\alpha,k} \; , \\
\nonumber
\delta v_I^x&=&\sum_{\alpha=1}^m\sum_k\rho_{\alpha}(k)\delta v_I^{x,\alpha,k} \; , \\
\delta v_I^y&=&\sum_{\alpha=1}^m\sum_k\rho_{\alpha}(k)\delta v_I^{y,\alpha,k} \; , \\
\nonumber
\delta a_{I_0}&=&\sum_{\alpha=1}^m\Biggl\{\rho_{\alpha}(0)\left[\delta a_{I_0}^{\alpha,0}+
\frac{1}{2}\sum_{J=1}^m A_J^{\alpha}(0)\frac{\partial\mu_I}{\partial \kappa_J}\right]\\
\nonumber
&&+\sum_{k\neq 0}\rho_{\alpha}(k)\left[\delta a_{I_0}^{\alpha,k}+\frac{1}{2} A_I^{\alpha}(k)
\frac{\omega_{\alpha}}{k}e^{-iky}\right]\Biggr\} \; ,\\
\nonumber
\delta\theta_I&=&\frac{1}{2}\sum_{\alpha=1}^m\left[\rho_{\alpha}(0) A_I^{\alpha}(0)y
+\sum_{k\neq 0}\rho_{\alpha}(k)\delta\theta_I^{\alpha,k}\right] \; .
\ea
In (\ref{expansion}) the time dependence has been shifted from the eigenmodes to the expansion
coefficients $\rho_{\alpha}(k)$ which are going to become upon quantization the creation and annihilation
operators of the edge excitations. Similarly the coefficients $\rho_{\alpha}(0)$ will play
the role of the number operators of the zero modes [the $k\rightarrow 0$ limit of the real part of
(\ref{lsol})]. They, in turn, are related to the number operators $\rho_J(0)$ which measure the total
excess charge at the edge relative to the background mean-field configuration due to the presence of
translation modes $(2\pi m/\nu L)(\partial\rho_I/\partial\kappa_J)$. Explicitly
\be
\label{relation}
\rho_I(0)=\frac{L\nu}{2\pi m}\sum_{\alpha=1}^m A_I^{\alpha}(0)\rho_{\alpha}(0) \; .
\ee
In the following we take these operators to be time independent corresponding to fixed
total charge at the edge.

The effective low energy Lagrangian is obtained by substituting (\ref{expansion}) into (\ref{lagrangian})
and keeping terms to lowest order in $k$ ,
\ba
\label{l2eq}
\nonumber
L_{sym}=&-&\frac{2\pi\hbar}{L\nu}\sum_{k>0}\frac{i}{k}\partial_t\rho_1(k)\rho_1(-k) \\
\nonumber
      &\mp&\frac{2\pi\hbar}{L}\sum_{\alpha=2}^m\sum_{k>0}\frac{i}{k}\partial_t\rho_{\alpha}(k)
          \rho_{\alpha}(-k) \\
\nonumber
     &-&\frac{\pi}{L\nu}[\eta_1+(m-1)\eta_2]\rho_1(0)\rho_1(0) \\
     &\mp&\frac{\pi}{L}\left(\eta_1-\eta_2\right)\sum_{\alpha=2}^m\rho_{\alpha}(0)\rho_{\alpha}(0) \\
\nonumber
     &-&\frac{2\pi\hbar}{L\nu}\sum_{k>0}v_1(k)\rho_1(k)\rho_1(-k) \\
\nonumber
     &\mp&\frac{2\pi\hbar v_2}{L}\sum_{\alpha=2}^m\sum_{k>0}\rho_{\alpha}(k)\rho_{\alpha}(-k) \; ,
\ea
where we have restored the units of dimensions.
The first two terms of the Lagrangian constitute its symplectic
part containing the information about the algebra satisfied by the density operators. Taking
$\rho_{\alpha}(k>0)$ to be the coordinates we find that $\rho_{\alpha}(k<0)$ (up to constants) play
the role of their conjugate momenta. Accordingly their commutation relations are given by ($k>0$)
\ba
\label{commuteq}
\nonumber [\rho_1(k),\rho_1(-p)]&=&-\frac{kL\nu}{2\pi}\delta_{kp} \; , \\
\nonumber [\rho_{\alpha}(k),\rho_{\beta}(-p)]&=&\mp\frac{kL}{2\pi}\delta_{\alpha\beta}
\delta_{kp}\; , \;\;\;\;\;\;\;\;\; \alpha,\beta=2,\cdots,m  \\
\, [\rho_1(k),\rho_{\alpha}(-p)]&=& 0 \; .
\ea
The number operators $\rho_{\alpha}(0)$, or equivalently $\rho_I(0)$, are absent from
the above algebra due to our assumption of constant edge charge.
In case this assumption does not hold one needs to introduce additional operators,
conjugated to $\rho_I(0)$, that change this charge\cite{vonDelft98}.

The effective low energy Hamiltonian is given by (minus) the remaining part of the Lagrangian.
Its first two terms correspond to the energy due to excess charge on the edge while the last
two terms describe the energy of the gapless excitations. Using the commutation relations
(\ref{commuteq}) one finds that $\rho_{\alpha}(k)$ indeed create and destroy the eigenmodes
(\ref{lsol}) with frequencies $\omega_{\alpha}(k)$. Note that for the Hamiltonian to be bounded
from below the velocity of the magnetoplasmon $v_1$ should be positive
while the neutral modes must propagate along the $\pm y$ direction ($v_2\gtrless 0$) for the
fractions $\nu=m/(pm\pm 1)$, as argued previously and affirmed by our numerical calculations.

To lowest order in $k$ we find that in the gauge defined by (\ref{landauaxy},\ref{landaua0})
the bosonic operators are given by $\phi_I({\bf r})=\sqrt{\rho_I(x)}\exp{[i(\theta_I+\delta
\theta_I)(y)]}$. Using Eqs. (\ref{f},\ref{g}) the CS transformations (\ref{cstransform1},
\ref{cstransform2}) can be inverted to obtain from $\phi_I({\bf r})$ an approximate form for the
electronic operator $\psi_I({\bf r})$. It contains the low energy edge components of the exact
$\psi_I({\bf r})$. As a final step we project the two-dimensional operator on the edge
by integrating over the $x$ direction with the weight $\partial_\kappa\rho_I(x)/\sqrt{\rho_I(x)}$.
The resulting expression for the one-dimensional operator [after omitting $y$-independent terms in
$\Lambda({\bf r})$ and $\Lambda_I({\bf r})$] is
\ba
\label{eqboson}
\nonumber \psi_I(y)&=&L^{-1/2}\exp[i\varphi_I(y)] \; ,\\
          \varphi_I(y)&=&\kappa y +\sum_{\alpha=1}^mA_I^{\alpha}(0)\rho_{\alpha}(0)y \\
\nonumber &+&\sum_{k\neq 0}\sum_{\alpha=1}^m\frac{i}{k}A_I^{\alpha}(k)\rho_{\alpha}(k)e^{-iky} \; ,
\ea
which constitutes the bosonization formula of the CTLL model.
It is interesting that in the gauge we are using $\varphi_I$
is composed in equal parts of contributions from the CS phase $\Lambda+\Lambda_I$ and from the
phase $\delta\theta_I$ of the bosonic field. We also note, using (\ref{relation},\ref{eqboson}),
that the spectrum of $\rho_I(0)$ is constrained by the periodic conditions on $\psi_I$ to take discrete
values in multiples of $\nu/m$. This observation conforms with our earlier identification of the
translation modes as edge states of quasiparticles [see discussion following (\ref{lbfield}-
\ref{lsch})].

The above expression for $\psi_I$ obeys fermionic anticommutation relations. It is a result of
\be
\label{eqphicomm}
[\varphi_I(y),\varphi_J(y')]=(p\pm\delta_{IJ})\pi i\;{\rm sgn}(y-y') \; ,
\ee
which is easily obtained using the orthogonality relation $\sum_{\alpha=2}^m(A_I^{\alpha})^2
=(2\pi/L)^2(1-1/m)$ (in order to have the proper anticommutation between different $\psi_I$
one also needs to introduce Klein factors\cite{vonDelft98}.)
Additionally Eqs. (\ref{expansion},\ref{commuteq}) can be
utilized to derive $[\delta\rho(y),\psi_I^{\dagger}(y')]=\delta(y-y')\psi_I^{\dagger}(y')$, where
$\delta\rho(y,t)=\int dx\sum_{I=1}^m\delta\rho_I({\bf r},t)$, indicating that $\psi_I^{\dagger}$
creates a localized unit charge at the edge as it should.

The bosonization formula (\ref{eqboson}) and the fact that the $\rho_\alpha$ create eigenmodes of
the system enable a straightforward evaluation of the edge propagator, which in the case of
short range interaction is
\be
\label{eqprop}
\langle\psi_I^{\dagger}(y,t)\psi_I(0,0)\rangle\propto (y-v_1 t)^{-\frac{1}{\nu}}(y- v_2 t)^
{-\left(1-\frac{1}{m}\right)} \; .
\ee
The theory thus reproduces the power-law behavior $I\sim V^{\gamma}$ for tunneling into
FQHE edges as predicted previously by Wen \cite{Wen95}. One finds $\gamma=p+1$ for the
$\nu=m/(pm+1)$ fractions and $\gamma=p+1-2/m$ if $\nu=m/(pm-1)$. In the presence of Coulomb interactions
the magnetoplasmon contribution to the propagator (\ref{eqprop}) introduces an additional correction
$\sim(\ln t)^{-\frac{1}{\nu}}$ at long times $2\nu\lambda_c \omega_c t\gg 1$.

So far we did not take into account the possibility of tunneling between the various condensates.
Such processes may exist in multi-layer systems with small inter-layer separation (of the kind needed
to establish correlations between the layers as assumed in our theory). They necessarily arise in
single-layer FQHE systems from the terms connecting different composite fermions Landau levels as discussed
following Eq. (\ref{lagrangian1}). In an $m=2$ symmetric edge Eq. (\ref{eqboson}) implies that
$\psi_{I=1,2}\propto\exp[i(\varphi_c\pm \varphi_n)]$ where $\varphi_c$ and $\varphi_n$ are the
magnetoplasmon and the neutral mode contributions to $\varphi_I$ respectively. Consequently a
tunneling term $\psi_1^\dagger\psi_2+{\rm h.c.}$ modifies the neutral sector of the free edge theory
(\ref{l2eq}) into a {\it chiral} sine-Gordon theory. The edge propagator for a $\nu=2/(2p+1)$ system
under such circumstances has been calculated by Naud et.al\cite{Naud00} and an algebraic decay,
with the same exponent as in Eq. (\ref{eqprop}), was found to persist.

\section{Asymmetrical Edges}
\label{sec:Asym}

When the edge configuration is asymmetric the matrix $B$ appearing in the secular equation (\ref{secular})
is no longer symmetric and the eigenvectors ${\bf A}^{\alpha}$ do not form in general an orthogonal set.
More notably the latter do not correspond to eigenmodes of the effective Hamiltonian and an additional
diagonalization procedure is needed.

To demonstrate this point we proceed to evaluate the effective Lagrangian, as described in the
previous section, continuing to use the normalization $\sum_{I=1}^m A_I^{\alpha}=2\pi
m/L\nu$ in which the solutions carry a unit charge. The symplectic part yields the following algebra
for the density operators
\ba
\label{commutneq}
\nonumber &&[\rho_{\alpha}(k),\rho_{\beta}(-p)]=-\frac{kL}{2\pi}\left(M^{-1}\right)_{\alpha\beta}
\delta_{kp} \;\;\;\;\;\;\;\;\;\; (k>0) \; , \\
&&M_{\alpha\beta}=\mp p(pm\pm 1)\pm\left(\frac{L}{2\pi}\right)^2\sum_{I=1}^m A_I^{\alpha}A_I^{\beta}
\; .
\ea
Note that now ${\bf A}^{\alpha}$ and therefore also $M$ are (even) functions of $k$. For notational
convenience we avoid an additional $k$ index on these quantities. The effective Hamiltonian reads
\ba
\label{neqh}
\nonumber H&=&\frac{L\hbar}{4\pi}\sum_{\stackrel{I,J,L}{\alpha ,\beta=1}}^m \frac{\partial\mu_I}
{\partial\kappa_J}\left(K^{-1}\right)_{IL} A_L^{\alpha}A_J^{\beta}\rho_{\alpha}(0)\rho_
{\beta}(0)\\
&+&\frac{\pi\hbar}{L}\sum_{\alpha, \beta=1}^m\sum_{k>0} M_{\alpha\beta}\frac{\omega_{\alpha}+\omega_
{\beta}}{k} \rho_{\alpha}(k)\rho_{\beta}(-k) \; .
\ea
It is now clear that the frequencies $\omega_{\alpha}$ obtained from the secular equation
are not the eigenfrequencies of $H$ and $\rho_{\alpha}$ do not create its eigenmodes. Nevertheless we can
still express the $\psi_I$ in terms of them. The result is similar to the one found in the symmetric case
(\ref{eqboson}) except that now $\kappa$ is replaced by $\kappa_I$. Using the identity proven in
the appendix one can show that (\ref{eqphicomm}) and consequently the anti-commutation relations
and the charge associated with $\psi_I$ remain unchanged with respect to the symmetric case.

In the following we will derive the eigenmodes of (\ref{neqh}), consider their coupling to
an external electric field and calculate the tunneling exponent into the edge. We will consider a
generic fraction of the form $\nu=m/(pm+1)$ and restrict most of the discussion of
the other fractions to $\nu=2/3$.

\subsection{$\nu=m/(pm+1)$}

In order to make progress we note that the matrices $M$ and $K^{-1}$ are congruent
\be
\label{cong}
M=(pm\pm 1)^2 A K^{-1} \widetilde{A} \; ,
\ee
where $A$ is defined in (\ref{Mbar,A}). Accordingly they have the same signature. Since the spectrum
of $K^{-1}$ is composed of the eigenvalue $\nu/m$ and $m-1$ degenerate states with eigenvalues $\pm 1$
we conclude that $M$ has $m$ positive eigenvalues for the $\nu=m/(pm+1)$ fractions and a single positive
and $m-1$ negative eigenvalues if $\nu=m/(pm-1)$.

Thus, there exists a (real nonorthogonal) matrix
$U$ such that $UM^{-1}\widetilde{U}=\sigma$ where $\sigma$ is diagonal containing the signs of the
eigenvalues of $M^{-1}$. Concentrating on the $\nu=m/(pm+1)$ case, where $\sigma={\bf 1}$, we use it to
define for $k>0$ the bosonic operators
\be
\label{a+}
a_{\alpha k}=\left(\frac{2\pi}{kL}\right)^{1/2}\sum_{\beta=1}^m \left(VU\right)_{\alpha\beta}
\rho_{\beta}(-k) \; ,
\ee
in terms of which
\be
\label{rho+}
\rho_{\alpha}(-k)=\left(\frac{kL}{2\pi}\right)^{1/2}\sum_{\beta=1}^m \left(U^{-1}\widetilde{V}
\right)_{\alpha\beta}a_{\beta k} \; .
\ee
$V$ is an orthogonal matrix to be determined below. The hermitian conjugates of (\ref{a+},\ref{rho+})
are obtained using $\rho_{\alpha}(-k)=\rho_{\alpha}^{\dagger}(k)$. Defining the real symmetric
matrices
\be
\label{NR}
N_{\alpha\beta}=\frac{1}{2}M_{\alpha\beta}(\omega_{\alpha}+\omega_{\beta}) \;\;\;\; , \;\;\;\;
R=\widetilde{U}^{-1}NU^{-1} \; ,
\ee
the $k\neq 0$ part of $H$ (\ref{neqh}) takes the form
\begin{eqnarray}
\label{diagh+}
&&\frac{2\pi\hbar}{L}\sum_{\alpha, \beta=1}^m\sum_{k>0} \frac{N_{\alpha\beta}}{k}\rho_{\alpha}(k)
\rho_{\beta}(-k)= \\
\nonumber
&&\hspace{4.5cm}\hbar\sum_{\alpha=1}^m\sum_{k>0}\Omega_{\alpha}(k)a_{\alpha k}^{\dagger}a_{\alpha k} \; ,
\end{eqnarray}
provided we choose $VR\widetilde{V}=\Omega$ with $\Omega$ diagonal. Thus, once $U$ and $V$ have been
calculated the eigenmodes of $H$ are determined. In practice, however, we find them to be very close
to the solutions of the secular equation (\ref{secular}). As a result, in the long wavelength limit and
in the presence of Coulomb interactions the distinction between a single magnetoplasmon
mode and $m-1$ modes with acoustic dispersion is preserved. The dispersion relations receive small,
linear in $k$, corrections which lift the degeneracy between the acoustic modes. For short range
interactions our numerical results indicate that increasing the amount of asymmetry tend to increase
the velocity of the in-phase mode while decreasing the velocity of the out-of-phase oscillations.

Expressing $\varphi_I$ through $a_{\alpha k}$ we can calculate the edge propagator. Assuming short
range interaction we obtain
\ba
\label{neqprop+}
&&\langle\psi_I^{\dagger}(y,t)\psi_I(0,0)\rangle\propto \prod_{\alpha=1}^m
(y-v_{\alpha} t)^{-\gamma_I^{\alpha}} \; ,\\
\nonumber
&&\gamma_I^{\alpha}=\left(\frac{L}{2\pi}\right)^2\sum_{\beta ,\gamma=1}^m A_I^{\beta}A_I^{\gamma}
\left(U^{-1}\widetilde{V}\right)_{\beta\alpha}\left(V\widetilde{U}^{-1}\right)_{\alpha\gamma} \; .
\end{eqnarray}
Using $U^{-1}\widetilde{U}^{-1}=M^{-1}$ and the result of the appendix one finds that the time
dependence of the propagator is insensitive to the details of the edge configuration
\begin{equation}
\label{gamma+}
\gamma_I=\sum_{\alpha=1}^m\gamma_I^{\alpha}=p+1 \; ,
\end{equation}
in agreement with the universal tunneling exponent predicted by Wen \cite{Wen95}.

\begin{figure}[ht!!!]
\setlength{\unitlength}{1in}
\psfig{figure=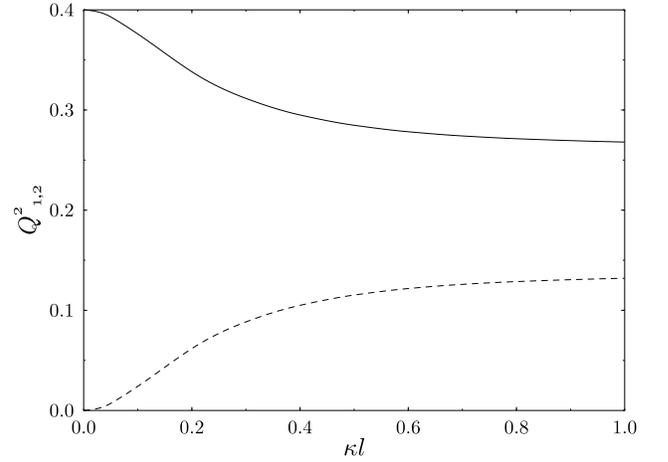,angle=270,width=3.3in}
\caption{Mode couplings for neutral asymmetric configurations $\kappa_1=-\kappa_2=\kappa$
of a $\nu=2/5$ edge with short range interaction $\lambda_s=1$ and $\lambda_{IJ}=2\pi m/\nu\delta_{IJ}$.
The solid and dashed lines correspond to $Q_1^2$ and $Q_2^2$ respectively.}
\label{q12+}
\end{figure}

For asymmetrical configurations the acoustic modes are no longer neutral. Consequently they couple
to an external potential. The response of the edge to such a potential is given by the retarded
density-density correlation function which is given by
\ba
\label{neqc+}
&&C_{\rm ret}(k,\omega)=-\sum_{\alpha=1}^m\frac{Q_{\alpha}^2(k)}{2\pi}
\frac{k}{{\omega-\Omega_{\alpha}(k)+i\delta}} \; , \\
\nonumber
&&Q_{\alpha}=\sum_{\beta=1}^m\left(V\widetilde{U}^{-1}\right)_{\alpha\beta} \; .
\ea
The mode couplings obey the sum rule \cite{Wen-papers}
\be
\label{sumrule+}
\sum_{\alpha=1}^m Q_{\alpha}^2=\sum_{\alpha, \beta=1}^m\left(M^{-1}\right)_{\alpha\beta}=
\nu \; ,
\ee
as shown in the appendix. Fig. \ref{q12+} depicts the way the couplings evolve as a $\nu=2/5$ edge
with short range interactions is taken out of equilibrium. Clearly the charge of the out-of-phase mode
increases with the degree of asymmetry, at the expense of the in-phase mode. In the case of Coulomb
interactions a similar trend exists but the couplings become $k$-dependent. In particular the
coupling of the acoustic mode decreases as $1/\ln(k)$ in the long-wavelength limit
(see also Fig. \ref{coupling2/3}) thus making it more difficult to detect.

\begin{figure}[ht!!!]
\setlength{\unitlength}{1in}
\psfig{figure=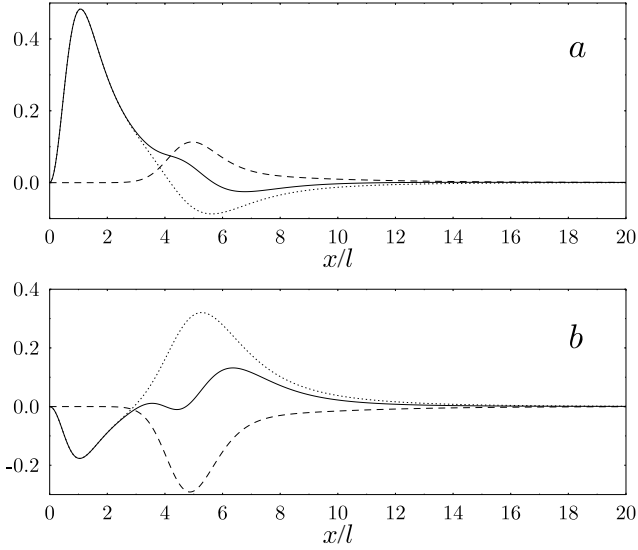,angle=270,width=3.3in}
\caption{Edge excitations of a neutral $\nu=2/3$ edge out of equilibrium $\kappa_1=-\kappa_2=2$
with Coulomb interaction $\lambda_c=0.2$ and $\lambda_{IJ}=2\pi m/\nu\delta_{IJ}$. They were calculated
assuming a wavevector $kl=10^{-4}$. The magnetoplasmon is shown in (a) where the dashed, dotted
and solid lines present the density variations  $\delta\rho_1^1$,$\delta\rho_2^1$ and their sum,
respectively. The corresponding quantities for the acoustic mode are shown in (b).}
\label{c2/3exc}
\end{figure}

\subsection{$\nu=2/3$}

Here the edge supports two modes with $\sigma_1=1$ and $\sigma_2=-1$. The introduction of the bosonic
operators
\ba
\label{a-}
a_{\alpha k}&=&\left(\frac{2\pi}{kL}\right)^{1/2}\sum_{\beta=1}^2 U_{\alpha\beta}
\rho_{\beta}(-\sigma_{\alpha}k) \; , \\
\nonumber
\rho_{\alpha}(-k)&=&\left(\frac{kL}{2\pi}\right)^{1/2}\left[\left(U^{-1}\right)_{\alpha 1}a_{1k}+
\left(U^{-1}\right)_{\alpha 2}a_{2k}^{\dagger}\right] \; ,
\ea
does not lead to a diagonalized $H$ and an additional Bogoliubov transformation
\be
\label{bogol}
b_{1,2 k}=\cosh\theta_k a_{1,2 k}-\sinh\theta_k a_{2,1 k}^{\dagger} \; ,
\ee
\begin{figure}[t!!!]
\setlength{\unitlength}{1in}
\psfig{figure=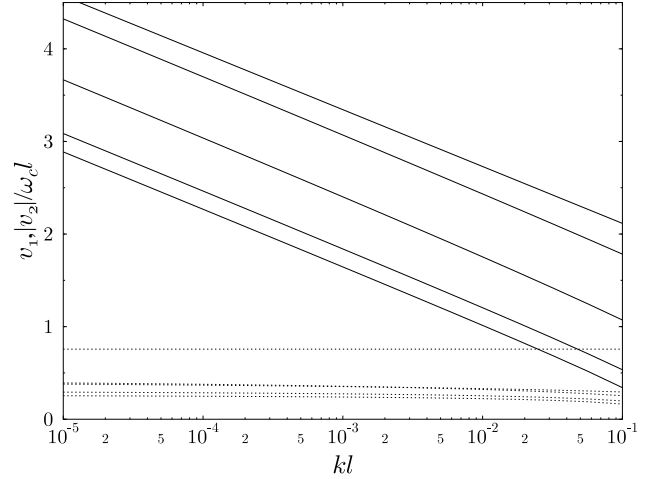,angle=270,width=3.3in}
\caption{Phase velocities of edge modes in neutral non-equilibrium configurations
$\kappa_1=-\kappa_2=\kappa$ of a $\nu=2/3$ edge with Coulomb interaction $\lambda_c=0.2$ and
$\lambda_{IJ}=2\pi m/\nu\delta_{IJ}$. The solid lines depict, from top to bottom $v_1$ for
$\kappa=0,1,2,3$ and 4. The dotted lines correspond in the same order to $|v_2|$.}
\label{fv12-}
\end{figure}
\vspace{1cm}

\noindent
with
\be
\label{thetak}
\tanh 2\theta_k=-\frac{2R_{12}}{R_{11}+R_{22}} \; ,
\ee
is needed to achieve the desired form. For an example of the resulting eigenmodes refer to Fig. \ref{c2/3exc}.

The eigenfrequencies are
\be
\label{omegas-}
\Omega_{1,2}(k)=\frac{1}{2}\left[\pm\left(R_{11}-R_{22}\right)+\sqrt{\left(R_{11}+R_{22}\right)^2
-4R_{12}^2}\, \right] \, .
\ee
While $\Omega_{1,2}$ are both positive Eqs. (\ref{a-},\ref{bogol}) imply that the two eigenmodes have
velocities of opposite signs. The mode with acoustic dispersion continues to propagate with negative
velocity $v_2=-\Omega_2/k$.

The velocities in the presence of Coulomb interaction are presented in Fig. \ref{fv12-}.
The parameter $\kappa$ controls the width of the outer strip in which the density is larger than its bulk
value $\nu=2/3$ (cf. Fig. \ref{fig-c2/3}). Varying its value introduces small $O(1/\ln k)$ corrections to the
logarithmic $k$ dependence of $v_1$ and to the constancy of $v_2$. It does, however, modify the $k$
independent parts of the velocities. Increasing the width decreases their values in accordance with the
findings of Ref. \onlinecite{Zulicke98a}. The numerical values found for $v_{1,2}$ are also quite close to
those obtained in Ref. \onlinecite{Zulicke98a}.

\begin{figure}[ht!!!]
\setlength{\unitlength}{1in}
\psfig{figure=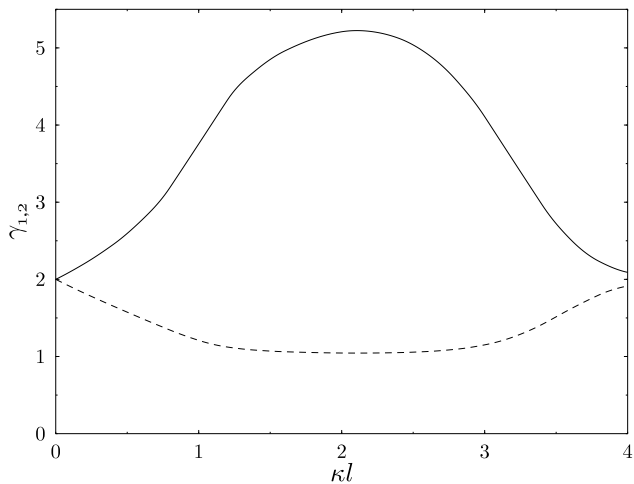,angle=270,width=3.3in}
\caption{The tunneling exponents $\gamma_1$ (solid line) and $\gamma_2$ (dotted line)
for neutral non-equilibrium configurations $\kappa_1=-\kappa_2=\kappa$ of a $\nu=2/3$ edge with short
range interaction $\lambda_s=0.2$ and $\lambda_{IJ}=2\pi m/\nu\delta_{IJ}$.}
\label{gamma12-}
\end{figure}

The edge propagator in the presence of short range interaction exhibits a power-law behavior
similar to (\ref{neqprop+}) but with non-universal exponents
\begin{widetext}
\be
\label{expo-}
\gamma_I=\sum_{\alpha=1}^2\gamma_I^{\alpha}=\left(\frac{L}{2\pi}\right)^2\sum_{\alpha, \beta=1}^2
A_I^{\alpha}A_I^{\beta}\left[\left(U^{-1}\widetilde{U}^{-1}\right)_{\alpha\beta}\cosh 2\theta_k
+2\left(U^{-1}\right)_{\alpha 1}\left(U^{-1}\right)_{\beta 2}\sinh 2\theta_k \right] \; .
\ee
\end{widetext}
In Fig. \ref{gamma12-} we present the tunneling exponents for neutral configurations with
$\kappa_1=-\kappa_2=\kappa$. Increasing $\kappa$ causes the leading edges of the two condensates to move
away from the wall. When the edge nearest to the wall (that of the $I=2$ condensate in this case)
looses contact with the wall we find that $\gamma_{1,2}$ recover their equilibrium values.

When tunneling occurs into the entire edge and not into a specific layer
the long time (low energy) behavior is determined by the smaller exponent
($\gamma_2$ in our case). Driving the system out of equilibrium has the effect of decreasing
this exponent. While this tendency agrees with the experimental findings \cite{Chang96,Grayson98,Chang00,Hilke01}
of a smaller exponent than predicted by the CTLL theory in tunneling into FQHE edges, it does not provide a
satisfactory explanations for the observed approximate $1/\nu$ dependence of the tunneling exponent over a
wide range of filling factors including those of the form $m/(pm+1)$. It should be noted here that in contrast
to the symmetric case it is still unclear whether inter-condensate tunneling in asymmetric edges may affect
the exponent.

Finally, the couplings of the modes to an electric perturbation along the edge, as deduced from
the density-density correlator (\ref{neqc+}) are
\be
\label{q-}
Q_{1,2}=\sum_{\alpha=1}^2\left(U^{-1}\right)_{\alpha \, 1,2}\cosh\theta_k+
\left(U^{-1}\right)_{\alpha \, 2,1}\sinh\theta_k \; .
\ee
\begin{figure}[ht!!!]
\setlength{\unitlength}{1in}
\psfig{figure=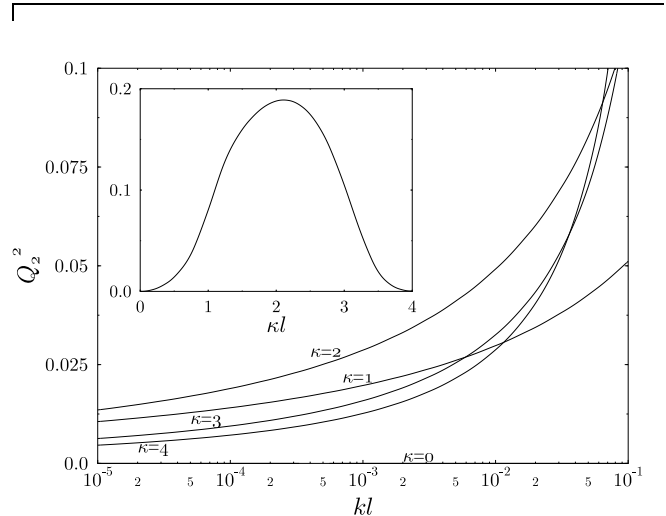,angle=270,width=3.3in}
\caption{The coupling of the acoustic mode for neutral asymmetric configurations
$\kappa_1=-\kappa_2=\kappa$ of a $\nu=2/3$ edge with Coulomb interaction $\lambda_c=0.2$
and $\lambda_{IJ}=2\pi m/\nu\delta_{IJ}$.
The inset presents $Q_2^2$ for similar configurations but in the presence of short range
interaction with $\lambda_s=0.2$. According to (\protect{\ref{sumrule-}}) the coupling of the
magnetoplasmon mode $Q_1^2$ is given by $2/3+Q_2^2$.}
\label{coupling2/3}
\end{figure}
They obey a similar sum rule to (\ref{sumrule+})
\be
\label{sumrule-}
\sum_{\alpha=1}^2 \sigma_{\alpha}Q_{\alpha}^2=\nu \; .
\ee
The presence of Coulomb interactions makes the couplings acquire a $1/\ln k$ dependence and reduces
the coupling of the acoustic mode relative to the case of short range interactions
(see Fig. \ref{coupling2/3}). For the latter type of interactions we observe that once $\kappa$ is
large enough such that the condensates are no longer in contact with the wall $Q^2_2$ vanishes.
The near neutrality of the counter-propagating mode makes its detection difficult especially in
experiments \cite{Ashoori92} where charge fluctuations are induced and detected over a wide range
near the edge. The increase of the coupling of this mode to an external probe once the edge is taken
out of equilibrium might help in its detection in similar experimental setups. Alternatively one
should excite the two condensates selectively by carefully placing a close top gate over part of
the edge region.

\section{CONCLUSION}

We have provided an explicit demonstration of the way in which the CTLL model for edges of
multi-layer FQHE systems emerges as the RPA low-energy limit of the CS theory for such systems.
The theory offers a unified description of edges both in and out of equilibrium and yields
an edge spectrum composed of a single magnetoplasmon branch and $m-1$ acoustic modes together
with their dependence on the underlying edge configuration. An attempt to carry the same
treatment over to single-layer FQHE states is possible only if one avoids reference to the
detailed structure of the composite fermion Landau levels and neglects transitions between
them. It is not clear to what extent such approximations are valid.

As far as the question of the tunneling exponent is concerned, we find agreement with the
universal results for the $\nu=m/(pm+1)$ states, obtained by previous studies. For $\nu=m/(pm-1)$
the minimal tunneling exponent depends on the structure of the edge and tends to
decrease when the edge is driven away from equilibrium. While this behavior may account for some
of the discrepancy between theory and experiment for $\nu>1/2$ it is inconsistent with the
observed approximate $1/\nu$ dependence of the exponent over a wide range $1>\nu>1/4$ of filling
factors containing both compressible and incompressible states. Moreover, additional processes,
not taken into account here such as impurity scattering, may drive the system back to equilibrium
and diminish the effect\cite{Kane95}. We hope that a careful study of
the approximations needed in order to apply the theory to single-layer systems, as mentioned
above, may help to settle this issue.

\acknowledgments
This research was supported by the United States - Israel Binational Science Foundation
(grant No. 2004162).

\appendix*
\section{}

In this appendix we prove the identity
\be
\label{iden}
\left(\frac{L}{2\pi}\right)^2\sum_{\alpha ,\beta=1}^m\left(M^{-1}\right)_{\alpha\beta}
A_I^{\alpha}A_J^{\beta}=p\pm\delta_{IJ} \; ,
\ee
used in calculating the commutation relations and the propagator of $\psi_I$. To this end we
define the matrices
\be
\label{Mbar,A}
\bar{M}_{\alpha\beta}=\frac{M_{\alpha\beta}}{(pm\pm 1)^2} \;\;\;\; ,\;\;\;\;
A_{\alpha I}=\frac{A_I^{\alpha}}{\sum_{I=1}^m A_I^{\alpha}} \; ,
\ee
in terms of which we can write (\ref{commutneq}) as
\be
\label{matM}
\bar{M}=\mp\frac{p}{pm\pm 1}C\pm A\widetilde{A} \; ,
\ee
where $C$ is the pseudo-identity matrix and tilde denotes transposition. Using the fact
$\sum_{I=1}^m A_{\alpha I}=1$ we have $AC=C$ and thus also $A^{-1}C=C$ and $C\widetilde{A}
^{-1}=C$. As a result
\be
\label{T}
T\equiv\pm A^{-1}\bar{M}\widetilde{A}^{-1}=1-\frac{p}{pm\pm 1}C \; .
\ee
Its inverse is readily found to be
\be
\label{invT}
T^{-1}=\pm\widetilde{A}\bar{M}^{-1}A=1\pm pC \; ,
\ee
which is just Eq. (\ref{iden}). Summing over $I$ and $J$ in Eq. (\ref{iden}) we obtain
\be
\label{msum}
\sum_{\alpha, \beta=1}^m\left(M^{-1}\right)_{\alpha\beta}=\nu \; ,
\ee
thus implying the sum rules (\ref{sumrule+},\ref{sumrule-}).


\begin{thebibliography}{999}
\bibitem{note2} There are other configurations which remain in the lowest Landau level but involve
topological defects, i.e. quasi-holes, that can not be reached by a smooth variation of the constant
solutions.
\bibitem{note3} The value of $\lambda_c$ used in calculating the static mean-field configuration
may be different than the one appropriate for determining the edge excitations. One should use
$\epsilon= \epsilon_{{\rm GaAs}}$ for both purposes as long as $kd>1$ where $k$ is the wavelength
of the excitation and $d$ the distance from the 2D electronic system to the surface of the device.
In most experimental cases the latter condition is not fulfilled and $\epsilon\simeq 1$ is a
reasonable approximation when considering the excitations, see Ref. \onlinecite{Volkov86}. We use
$\epsilon=(1+\epsilon_{{\rm GaAs}})/2$ in both calculations which results in $\lambda_c=0.2$ for a
typical $\nu=2/3$ FQHE sample.
\bibitem{note4} In the uniform system a quasiparticle of type $I$, localized at the origin, is
described by a mean-field solution with $\theta_{J}=\varphi\delta_{IJ}$ where $\varphi$ is the
azimuthal angle. The excess charge carried by this solution is
$\sum_{J=1}^{m}\left(K^{-1}\right)_{IJ}=\nu/m$.
\end{thebibliography}
\end{document}